# Controlled islanding for a hybrid AC/DC grid with VSC-HVDC using semi-supervised spectral clustering

Yang Li [a,b], Shangsong Wu [a]

[a] School of Electrical Engineering, Northeast Electric Power University, Jilin 132012, PR China

[b] Energy Systems Division, Argonne National Laboratory, IL 60439, United States

Corresponding author: Yang Li (e-mail: liyang@neepu.edu.cn).

**ABSTRACT** As the last resort of emergency control, controlled islanding is an effective means of preventing fault-propagation and a system-wide blackout. However, conventional AC transmission lines are unavailable to be employed for power exchange between these islands. To make full use of the DC power modulation capability of VSC-HVDC links, a new controlled islanding model is put forward for an AC/VSC-HVDC hybrid grid to minimize the composite power-flow disruption, in which the DC-terminals belonging to a VSC-HVDC link are placed in different islands. To solve this model, a semi-supervised spectral clustering-based approach is proposed by transforming the problem into a weighted undirected graph segmentation problem. The novelty of our work is to find an optimal islanding solution in real time such that the power exchanges between islands are implemented via a VSC-HVDC link to reduce the generation-load imbalance. The simulation results on the IEEE 39-bus system and a real-world system verify the effectiveness and superiority of the proposed approach.

## NOMENCLATURE

**Acronyms**

| | |
|---|---|
| AC | Alternating Current |
| DC | Direct Current |
| VSC | Voltage source converter |
| VSC-HVDC | Voltage source converter based high voltage direct current |
| NP-hard | Non-deterministic polynomial-time hard |
| OBDD | Ordered binary decision diagram |
| WAMS | Wide area measurement system |
| GPS | Global positioning system |
| SCADA | Supervisory control and data acquisition |
| Ncut | Normalized cut |
| SOM | Self-organizing maps |

**Symbols**

| | |
|---|---|
| $R$ | Resistance between the AC and DC network |
| $X$ | Reactance between the AC and DC network |
| $P_s$ | Active power injected from the AC network |
| $Q_s$ | Reactive power injected from the AC network |
| $P_c$ | Active power injected from the AC grid into VSC |
| $Q_c$ | Reactive power injected from the AC grid into VSC |
| $U_s$ | AC bus voltage |
| $U_c$ | Converter voltage |
| $U_{dc}$ | Voltage of DC grid |
| $D_{ij}$ | Electrical distance between bus $i$ and $j$ |
| $Z_{in}$ | Input impedance of a two-port network |
| $Z_{ii}$ | Self-impedances of bus $i$ |
| $Z_{jj}$ | Self-impedances of bus $j$ |
| $Z_{ij}$ | Mutual impedance between bus $i$ and $j$ |
| $p_{ij}$ | Active power between bus $i$ and $j$ |
| $P_{ij}$ | Power-flow disruption between bus $i$ and $j$ |
| $k$ | Number of clusters |
| $S$ | Power system |
| $n$ | Buses |
| $L$ | Lines |
| $N_{Gen}$ | Generator |
| $S_u$ | Subsystem of coherent group $u$ |
| $S_t$ | Subsystem of coherent group $t$ |
| $\tilde{P}$ | Minimum composite power-flow disruption |
| $\Delta P_i$ | Active power imbalance of bus $i$ |
| $\Delta Q_i$ | Reactive power imbalance of bus $i$ |
| $U_i$ | Voltage amplitudes of buses $i$ |
| $U_j$ | Voltage amplitudes of buses $j$ |
| $G_{ij}$ | Conductance between bus $i$ and $j$ |
| $B_{ij}$ | Susceptance between bus $i$ and $j$ |
| $\delta_{ij}$ | Phase difference between $U_i$ and $U_j$ |
| $G$ | Conductance between the AC and DC network |
| $B$ | Susceptance between the AC and DC network |
| $P_{Gen}$ | Generator active power |
| $Q_{Gen}$ | Generator reactive power |
| $I_{dc}$ | Current of DC grid |
| $S_0$ | Center of the PQ-capability circle |
| $B_{Gen,i}$ | Generator buses $i$ |
| $B_{Gen,j}$ | Generator buses $j$ |
| $L_{ij}$ | Line connecting buses $i$ and $j$ |
| $B_{VSC1}$ | Terminal buses $VSC1$ |
| $B_{VSC2}$ | Terminal buses $VSC2$ |
| $G_0$ | Weighted undirected graph |
| $V$ | Point set of graph $G$ |
| $E$ | Edge set of graph $G$ |
| $W$ | Weighted adjacency matrix of graph $G$ |
| $w_{ij}$ | Weight value of the edge $(i, j)$ |
| $c$ | Cut |
| $A$ | Degree matrix |
| $a$ | Diagonal element of Degree matrix $A$ |
| $vol$ | Sum of degrees |
| $h$ | Indicator vector |
| $H$ | Indicator vectors matrix |
| $L$ | Non-normalized Laplacian matrix |
| $L_N$ | Normalized Laplacian matrix |
| $Tr$ | Trace of matrix |
| $E_0$ | Identity matrix |
| $x_i$ | $i$th data point |
| $\mu_{k'}$ | Center of cluster $k'$ |
| $v$ | Eigenvector |
| $\lambda$ | Eigenvalue |
| $M$ | Eigenvector matrix |
| $Rv$ | Row vector of $M$ |

**Superscripts**

| | |
|---|---|
| min | Lower limits |
| max | Upper limits |

## I. INTRODUCTION

Controlled islanding has always been regarded as one of the most important control measures to prevent cascading outages [1] or even large scope blackouts [2], and its key idea is to find optimal islanding solution with the minimum splitting costs while satisfying a group of various constraints related to system operation [3]. Recent research has demonstrated that the interconnection of large-scale power systems produces many well-known economic, social and environmental benefits, but on the other hand, it also significantly exaggerates the potential impact of a large disturbance [4, 5]. In this sense, the uncertainties in power system operation significantly increase under the growing penetration of new components such as distributed generations [6, 7, 8] and plug-in electric vehicles [9, 10], which inevitably leads to more serious consequences resulted from the loss of stability [11]. In addition, as a typical cyber-physical system, the increasing cyber-attacks, natural disasters and the reliance on control and communication are resulting in new sources and propagation paths of cascading failures in power systems [12, 13]. These shifts present new challenges in maintaining the system operating reliably and seamlessly. Unfortunately, available statistics indicate that major worldwide blackouts sharply increase from the 1960s [14], which have caused huge economic losses and harmful social influences. In this context, controlled islanding, as the last resort for preventing cascading outages, has received ever-increasing attention through splitting the system into several sustainable islands in the past few years [15, 16].

By monitoring of power system operating conditions, controlled islanding is to seek an optimal islanding solution quickly, so as to prevent faults from spreading and thereby leading to a system-wide blackout [17]. In essence, it is a typical non-deterministic polynomial-time hard (NP-hard) combinatorial optimization problem [1], and the so-called "combination explosion" in solution space significantly exacerbates the problem-solving difficulties with the increase of system size. Therefore, it is an urgent and challenging task to develop a computationally efficient controlled islanding approach to prevent cascading outages.

### A. LITERATURE REVIEW

Numerous studies have been carried out to solve controlled islanding problems such as graph partitioning-based method [18], slow coherency-based method [19-21], ordered binary decision diagram (OBDD)-based approach [1, 22, 23], $k$-means clustering [24], and self-organizing maps (SOM) neural networks [25]. The traditional graph partitioning technique is a widely used solution approach. In [18], both real and reactive power balances are considered to avoid the problem of low voltages of the isolated system due to insufficient reactive powers. However, the whole computation time of this method is too long because it is necessary to carry out the partition search calculation of the graph many times. In [19], the slow coherence theory is originally proposed for solving controlled islanding problems, and then a demonstration of this approach on the blackout scenario of August 14, 2003 in North America is given in [20]. Reference [21] presents a cutset determination algorithm based on slow coherency for controlled islanding of large power systems. The OBDD-based method is another effective method which solves the controlled islanding problem by reducing network of the original grid [22]. This method assumes that all lines of the system may be disconnected, and the complexity of the algorithm increases geometrically as the system scale increases. Therefore, before searching for a separation interface, the original network requires to be simplified to reduce search spaces and improve search efficiency [23]. However, the equivalence and simplification of the original network by these methods usually reduces solution space, so the obtained solution may not be the globally optimal solution [26]. In [24], a *k*-means clustering-based controlled islanding algorithm is presented in which VSC-HVDC links are employed to reduce load shedding. In [25], self-organizing maps (SOM) neural networks have been utilized in the defensive islanding. Recent research indicates that the graph theory-based spectral clustering technique is an effective solution approach for handling this controlled islanding problem by transforming the original issue into an optimal graph partition problem [26-28]. However, the traditional spectral clustering suffers some open problems such as the lack of ability to consider multiple constraints, which limits its usefulness in practical applications [29]. Recently, a two-step spectral clustering algorithm for controlled islanding has been proposed in [26]. However, if the number of islands is greater than 2, this problem must be solved via the time-consuming "recursive bisection" approach due to the repeated eigen-decomposition of a matrix in this process [27].

As a new HVDC transmission technology, VSC-HVDC is becoming popular in recent years [30, 31]. Compared with the traditional HVDC, VSC-HVDC has obvious advantages in regulating active and reactive power independently, supplying power for passive networks, interconnecting weak AC systems, etc [32-34]. Furthermore, a VSC-HVDC link is capable of exchanging powers between asynchronous islands, which helps to reduce the splitting costs, e.g., the generator tripping and load shedding, and accelerate the restoration process [24, 35]. Meanwhile, recent progress in a wide-area measurement system (WAMS) makes wide-area real-time synchronized measurements available for use, which opens up a new opportunity for developing a wide-area protection and control (WAPaC) system [36-38].

For a long time, there is no solution to the problem of power exchanges between islands, since it is impossible to do so by using AC transmission lines. The emergence of the VSC-HVDC technique breaks this technical bottleneck. With the wide application of VSC-HVDC and the unparalleled advantages in connecting isolated islands, the work on the controlled islanding problem for a hybrid AC/DC grid with VSC-HVDC is becoming very important.

The idea of connecting two isolated islands via a VSC-HVDC link has been proposed in [24, 35]. The distance represents the similarity between the nodes. The distance matrix is formed by modifying the distance of the nodes, and then the distance matrix is clustered by the *k*-means algorithm to obtain the optimal islanding scheme. Inspired by this work, a new controlled islanding approach for a hybrid AC/VSC-HVDC grid is presented using semi-supervised spectral clustering (SSSC), in which the power exchanges between islands are implemented via a VSC-HVDC link to reduce the generation-load imbalance.

Regarding controlled islanding, the coherence constraints must be guaranteed to ensure the stability of the post-split subsystems. Furthermore, for a hybrid AC/VSC-HVDC grid, the VSC-HVDC terminal constraints are of prime importance to achieve power exchanges between islands. As an extension of conventional spectral clustering algorithms, SSSC can ensure the boundaries of the clusters to be more correct by utilizing constraints as the prior information to guide the clustering process. For this reason, the SSSC is adopted for addressing the islanding system control in this study.

### *B. CONTRIBUTION OF THIS PAPER*

The contributions of this work include the following aspects.
(1) Using VSC-HVDC links to exchange powers between islands, a controlled islanding model is presented for a hybrid AC/VSC-HVDC grid. To reflect the electrical relation among buses, a new criterion is proposed as the objective function by combining electrical distance with the conventional minimal power-flow disruption. More importantly, the terminals of a VSC-HVDC link are placed in different islands in this model. By doing so, the power exchanges are implemented via VSC-HVDC links to reduce the generation-load imbalance.
(2) A new SSSC-based solution method is proposed, in which the coherence and VSC-HVDC terminal constraints are treated as the pair-wise constraints. In this way, both terminals of a VSC-HVDC link are ensured to be placed in different islands, while coherent generators are allocated into the same island.
(3) The test results on two test systems show that the proposed approach is capable to find the optimal islanding scheme where different islands are connected via a VSC-HVDC link to reduce the generation-load imbalance. In addition, our approach has proven to be superior to other state-of-the-art methods, such as the OBDD, SOM and *k*-means clustering.

### *C. ORGANIZATION OF THIS PAPER*

This paper is organized as follow. In Section II, VSC-HVDC and its control methods are introduced. Section III shows the problem formulation of controlled islanding for an AC/VSC-HVDC grid. Section IV describes the proposed approach in detail. In Section V, case studies are presented. And finally, conclusions are drawn in Section VI.

## II. VSC-HVDC AND CONTROL METHODS

Based on voltage source converters (VSCs) and the pulse-width modulation (PWM) technology, the emerging VSC-HVDC technology has some significant advantages such as flexible and rapid DC power modulation capability, which is ideal for interconnection of asynchronous islands [39]. From this point of view, this work focus on utilizing VSC-HVDC links to connect different islands. Fig. 1 shows a simplified model of an AC/VSC-HVDC system.

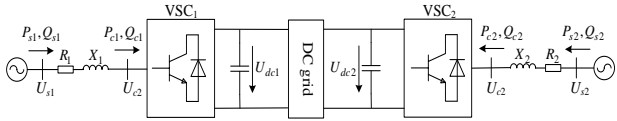

**FIGURE 1.** Simplified equivalent model of a two-terminal VSC-HVDC system

where $R_i$ ($i$=1, 2) and $X_i$ ($i$=1, 2) are respectively the resistance and reactance between the AC and DC network; $P_{si}$ ($i$=1, 2) and $Q_{si}$ ($i$=1, 2) denote the active and reactive power injected from the AC network, respectively; $P_{ci}$ ($i$=1, 2) and $Q_{ci}$ ($i$=1, 2) denote the active and reactive power injected from the AC grid into VSCs, respectively; $U_{si}$ ($i$=1, 2) represents the AC bus voltage; $U_{ci}$ ($i$=1, 2) represents the converter voltage, and $U_{dci}$ ($i$=1, 2) is the DC grid voltage.

The basic VSC-HVDC control methods include four categories [30, 40]: 1) constant DC voltage and AC voltage control; 2) constant active power and AC voltage control; 3) constant active and reactive power control; 4) constant DC voltage and reactive power control. Without loss of generality, the control strategy used in VSC$_1$ is the constant DC voltage and reactive power control, while that of VSC$_2$ is the constant active and reactive power control [30].

## III. PROBLEM FORMULATION

As demonstrated in [26-28], controlled islanding can be regarded as a clustering issue, in which the system is to be split into two or more clusters. In this work, a new criterion is put forward as the objective function by combining electrical distance with the conventional minimal power-flow disruption. Regarding constraints, the terminals of a VSC-HVDC link are constrained in different islands such that power exchanges between islands can be implemented via the VSC-HVDC link, besides conventional constraints.

### A. OBJECTIVE FUNCTION

When designing a controlled islanding algorithm, the used objective function is usually taken as the minimal power imbalance or the minimal power-flow disruption, which respectively refer to the minimal algebraic or the arithmetical sum of active powers on a separation interface. Compared with the minimal power imbalance, the use of minimal power-flow disruption can decrease the time complexity and enable it to be addressed effectively [26, 27]. On the other hand, recent research findings suggest that electrical distance plays an important role in the determination of the optimal islanding solution [41]. Considering the above two factors, a criterion is put forward as the objective function by combining the power-flow disruption and electrical distance.

The electrical distance between two buses $D_{ij}$ is defined as

$$D_{ij} = Z_{in} = Z_{ii} + Z_{jj} - 2Z_{ij} \quad (i \neq j) \tag{1}$$

where $Z_{in}$ denotes the input impedance of a two-port network, $Z_{ii}$ and $Z_{jj}$ are the self-impedances of bus $i$ and $j$, and $Z_{ij}$ is the mutual impedance of bus $i$ and $j$.

The power-flow disruption $|P_{ij}|$ can be expressed as [26, 27]

$$|P_{ij}| = (|p_{ij}| + |p_{ji}|) / 2 \tag{2}$$

where $p_{ij}$ denotes the value of the active power on the line between bus $i$ and $j$.

A power system $S$ with $n$ buses, $L$ lines, and $N_{Gen}$ generators is subjected to a large disturbance, and thereby this system is split into $k$ synchronous islands $S_1,...,S_u,...,S_t,...,S_k \subset S$ according to coherent generators. As the objective function, the new criteria is proposed as the for assessing splitting sections, which can be formulated as

$$\tilde{P} = \min_{S_1,S_2,\cdots,S_k \subset S} \left( \frac{1}{2} \sum_{i \in S_u, j \in S_t} \frac{|P_{ij}|}{D_{ij}} \right) (u,t = 1,2,\cdots,k, u \neq t) \tag{3}$$

where $\tilde{P}$ is the minimal composite power-flow disruption.

### B. CONSTRAINTS

The constraints mainly include three-folds: conventional constraints, coherence constraints, and VSC-HVDC terminal constraints.

1) CONVENTIONAL CONSTRAINTS

1.1) EQUALITY CONSTRAINTS

1.1.1) CONSTRAINTS OF AC NETWORK

The equality constraints of AC network are given by

$$\begin{cases} \Delta P_i = U_i \sum_{j=1}^{n} U_j \left( G_{ij} \cos \delta_{ij} + B_{ij} \sin \delta_{ij} \right) \\ \Delta Q_i = U_i \sum_{j=1}^{n} U_j \left( G_{ij} \sin \delta_{ij} - B_{ij} \cos \delta_{ij} \right) \end{cases} \quad (4)$$

where $\Delta P_i$ and $\Delta Q_i$ represent active and reactive power imbalance of bus $i$, respectively; $U_i$ and $U_j$ represent voltage amplitudes of buses $i$ and $j$, respectively; $G_{ij}$ and $B_{ij}$ represent conductance and susceptance between buses $i$ and $j$, respectively, and $\delta_{ij}$ denotes the phase difference of $U_i$ and $U_j$.

### 1.1.2) CONSTRAINTS OF DC NETWORK

The equality constraints of DC network are

$$\begin{cases} P_c = U_c^2 G - U_s U_c \left[ G \cos(\delta_s - \delta_c) + B \sin(\delta_s - \delta_c) \right] \\ Q_c = -U_c^2 B + U_s U_c \left[ G \sin(\delta_s - \delta_c) + B \cos(\delta_s - \delta_c) \right] \end{cases} \quad (5)$$

where $P_c$ and $Q_c$ denote active power and the reactive power injected from the AC grid into VSC, respectively; $G$ and $B$ represent conductance and susceptance between the AC and DC network, respectively; $\delta_s$ and $\delta_c$ represent phase angle of $U_s$ and $U_c$, respectively.

### 1.2) INEQUALITY CONSTRAINTS

### 1.2.1) CONSTRAINTS OF AC NETWORK

The inequality constraints of the AC network are

$$\begin{cases} P_{Gen,i}^{\min} \leq P_{Gen,i} \leq P_{Gen,i}^{\max}, & i=1,\ldots,N_{Gen} \\ Q_{Gen,i}^{\min} \leq Q_{Gen,i} \leq Q_{Gen,i}^{\max}, & i=1,\ldots,N_{Gen} \\ U_i^{\min} \leq U_i \leq U_i^{\max}, & i=1,\ldots,n \end{cases} \quad (6)$$

where $P_{Gen}^{\min}$ and $P_{Gen}^{\max}$ represent the lower and upper limits of generator active power $P_{Gen}$, respectively; $Q_{Gen}^{\min}$ and $Q_{Gen}^{\max}$ represent the lower and upper limits of generator reactive power $Q_{Gen}$, respectively; $U^{\min}$ and $U^{\max}$ represent the lower and upper limits of bus voltage $U$, respectively.

### 1.2.2) CONSTRAINTS OF DC NETWORK

The inequality constraints of the DC network are

$$\begin{cases} P_s^{\min} \leq P_s \leq P_s^{\max} \\ Q_s^{\min} \leq Q_s \leq Q_s^{\max} \\ U_{dc}^{\min} \leq U_{dc} \leq U_{dc}^{\max} \\ I_{dc}^{\min} \leq I_{dc} \leq I_{dc}^{\max} \\ r_{\min}^2 \leq (P_s - P_0)^2 + (Q_s - Q_0)^2 \leq r_{\max}^2 \end{cases} \quad (7)$$

where $P_s^{\min}$ and $P_s^{\max}$ are the lower and upper limits of the active power $P_s$ of DC grid, respectively; $Q_s^{\min}$ and $Q_s^{\max}$ represents the lower and upper limits of the reactive power $Q_s$ of DC grid, respectively; $U_{dc}^{\min}$ and $U_{dc}^{\max}$ represents the lower and upper limits of the voltage $U_{dc}$ of DC grid, respectively; $I_{dc}^{\min}$ and $I_{dc}^{\max}$ represents the lower and upper limits of the current $I_{dc}$ of DC grid, respectively; $S_0(P_0, Q_0)$ represents the center of the PQ-capability circle; $r_{\max}$ and $r_{\min}$ are the lower and upper limits of the radius $r$ of power circle, respectively.

### 2) COHERENCE CONSTRAINTS

Research findings in [1, 19, 26] show that coherence constraints that ensure coherent generators are divided in the same island are very important in addressing the controlled island problem, which is formulated as

$$\begin{cases} \forall B_{Gen,i}, B_{Gen,j} \in S_u, \exists L_{ij} \subset \Pi(B_{Gen,i} \cap B_{Gen,j}) \ (i \neq j) \\ \forall B_{Gen,i} \in S_u, B_{Gen,j} \in S_t, \Pi(B_{Gen,i} \cap B_{Gen,j}) = \varnothing \ (u \neq t) \end{cases} \quad (8)$$

where $B_{Gen,i}$ and $B_{Gen,j}$ represents generator buses $i$ and $j$, $L_{ij}$ indicates a line connecting buses $i$ and $j$. This equation indicates that there is at least one interconnection between any two coherent generators, and there is no interconnection to connect any two non-coherent generators.

### 3) VSC-HVDC TERMINAL CONSTRAINTS

To exchange powers between islands via a VSC-HVDC link, the terminals belonging to a VSC-HVDC link need to be located in different islands such that the generation-load imbalance in each island and the load shedding can be reduced. For this purpose, the VSC-HVDC terminal constraints can be formulated by

$$B_{vsc1} \in S_u, B_{vsc2} \in S_t, \Pi(B_{vsc1} \cap B_{vsc2}) = \varnothing \ (u \neq t) \quad (9)$$

where $B_{vsc1}$ and $B_{vsc2}$ represents two terminal buses $vsc1$ and $vsc2$ of a VSC-HVDC link.

## IV. PROPOSED SOLUTION ALGORITHM

A new SSSC-based method is proposed to solve the controlled islanding model for the AC/DC grid. The coherence constraints and VSC-HVDC terminal constraints are considered as pair-wise constraints, which generalize the traditional spectral clustering to SSSC to make the clusters more accurate. Meanwhile, the controlled islanding problem is transformed into a weighted undirected graph segmentation problem which can be solved efficiently via spectral clustering.

### A. SPECTRAL CLUSTERING ALGORITHM

#### 1) NORMALIZED CUT CRITERION

The system to be split can be modelled as a weighted undirected graph $G_0(V, E)$, where the point set $V$ contains all buses and the edge set $E$ contains all lines. The matrix $W$ is a weighted adjacency matrix of the graph $G_0$ and its element $W_{ij}$ is given by

$$W_{ij} = \begin{cases} w_{ij}, & (i,j) \in E \\ 0, & (i,j) \notin E \end{cases} \quad (10)$$

where $w_{ij}$ is the weight of edge $(i, j)$, $w_{ij} = w_{ji}$ since the matrix $W$ is symmetrical. In this study, $w_{ij} = |P_{ij}|/D_{ij}$. The higher the degree of similarity between the two points is, the greater the weight value is, and vice versa. If $W_{ij} = 0$, there is no connection between the two points. Here $A$ and $B$ are two disjoint complementary subsets of the graph $G_0$ ($A \cap B = \emptyset, A \cup B = V$). The *cut* between $A$ and $B$ is defined as:

$$c(A,B) = \sum_{i \in A, j \in B} w_{ij} \quad (11)$$

Extending to a general case, the sum of the *cut* among subsets can be expressed as:

$$c_{sum} = \frac{1}{2} \sum c(V_s, V_t) \quad (s,t=1,2,\ldots,k, s \neq t) \quad (12)$$

where $V_s$ and $V_t$ represent the disjoint subsets of $V$.

Thus, controlled islanding can be transformed into a graph segmentation problem. In order to make the sum of weight values in the same subset as large as possible, while the sum in different subsets is as small as possible, the traditional clustering method is to get the minimum cut directly. However, when this method encounters some special cases such as Fig. 2, it inevitably produces an isolated bus in a subset. In this study, the minimum normalized cut (*Ncut*) is utilized to solve the problem.

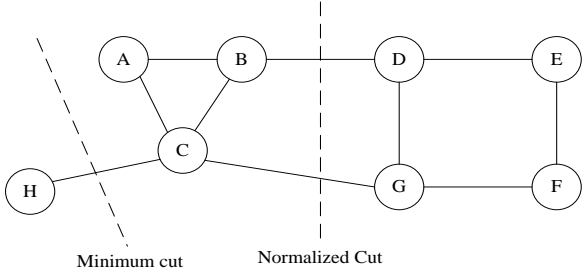

**FIGURE 2.** Comparison between minimum cut and normalized cut

Suppose matrix $A$ is a diagonal matrix and its diagonal elements are $a_{ii} = \sum_j w_{ij}$, the normalized cut of graph $G_0$ is

$$Ncut = \frac{1}{2} \sum \frac{c(V_s, V_t)}{vol(V_s)} \quad (s,t=1,2,\ldots k, s \neq t) \quad (13)$$

where $vol(V_s) = \sum_{i \in V_s} a_{ii}$ denotes the sum of the degrees of vertices in $V_s$.

In order to obtain the minimum normalized cut, $k$ cluster indicator vectors are defined as $h_j = (h_1, h_2, \ldots, h_n)^T$ ($j=1,2,\ldots k$) where

$$h_{ij} = \begin{cases} 1/\sqrt{vol(V_j)}, & i \in V_j \\ 0, & i \notin V_j \end{cases} \quad (i=1,2,\ldots,n; j=1,2,\ldots,k) \quad (14)$$

After derivation, the following equation can be obtained

$$h_j^T L h_j = 2 \frac{c(V_s, V_t)}{vol(V_s)} = 2Ncut \quad (t=1,2,\ldots,k, t \neq s) \quad (15)$$

where $L = A - W$, which denotes the non-normalized Laplacian matrix.

Suppose matrix $H = [h_1, h_2, \ldots, h_k] \in R^{n \times k}$, the minimum normalized cut is equivalent to

$$\begin{aligned} \min_{G_1, G_2, \ldots G_k} & \quad Tr(H^T L_N H) \\ s.t. & \quad H^T L_N H = E_0 \end{aligned} \quad (16)$$

where $Tr(H^T L_N H)$ is the trace of $H^T L_N H$, $L_N = A^{-1}(A-W) = A^{-1}L$ is the normalized Laplacian matrix and $E_0$ is an identity matrix.

For Eq. (16), the goal is to find the top $k$ minimum eigenvalues of the normalized Laplace matrix $L_N$ so that the corresponding $k$ eigenvectors can be obtained. According to the Rayleigh-Ritz method, $H$ is the matrix composed of eigenvectors corresponding to the top $k$ minimum eigenvalues. Considering the discrete combinatorial optimization characteristics in the solution process, the relaxation method is employed to relax matrix $H$ to be a real-valued matrix. After clustering each row of matrix $H$ via $k$-means algorithm, the final islanding solution can be obtained [29].

2) *k*-MEANS CLUSTERING

As one of the mostly used clustering algorithms, the $k$-means clustering is a vector quantization method, originally from signal processing, which seeks to the minimum within groups sum of squared errors (WGSS) through an iterative optimization process [42]. In general, this process is modeled as the following optimization problem.

$$\begin{aligned} \text{Minimize} & \quad \sum_{k'=1}^{k} \sum_{i=1}^{n} d_{k'i} \|x_i - \mu_{k'}\|^2 \\ s.t. & \quad \sum_{k'=1}^{k} d_{k'i} = 1, 1 \leq i \leq n \\ & \quad d_{k'i} \in \{0,1\}, 1 \leq i \leq n, 1 \leq k' \leq k \end{aligned} \quad (17)$$

where $x_i$ is the $i$th data point, $\mu_{k'}$ is the center of cluster $k'$, $\|x_i - \mu_{k'}\|^2$ denotes the squared Euclidean distance between $x_i$ and $\mu_{k'}$.

The specific procedures of the $k$-means clustering are as follows.

Step 1: Determine the number of clusters $k$, and select $k$ cluster centers randomly at the first iteration.

Step 2: Calculate the Euclidean distance between data

points and cluster centers, and thereafter the nearest data are divided into the same cluster.

Step 3: Update the $k$ cluster centers and the cluster sets.

Step 4: Determine whether the cluster centers in successive iterations are changed. If changed, output the final clustering results; otherwise, go to step 2.

### B. SEMI-SUPERVISED SPECTRAL CLUSTERING

The SSSC is an extension of the traditional spectral clustering, and its basic principle is to enable the boundaries of the clusters to be more correct with the use of prior information based on pair-wise constraints [29]. In this study, the pair-wise constraints refer to the Must-link constraint and the Cannot-link constraint: the former indicates that coherent generators must be clustered in the same group, while the latter means that non-coherent generators must be separated in different clusters. In order to consider the pair-wise constraints of generators, the weighted adjacency matrix $W$ needs to be modified as follows:

$$W\left(B_{Gen,i}, B_{Gen,j}\right) = \begin{cases} \infty, & \left(B_{Gen,i}, B_{Gen,j}\right) \in \text{Must-link} \\ 0, & \left(B_{Gen,i}, B_{Gen,j}\right) \in \text{Cannot-link} \end{cases} \quad (18)$$

From Eq. (18), it can be seen that these constraints strengthen the electrical contacts between coherent generators, and weaken the contacts between non-coherent generators.

Since VSC-HVDC terminal constraints belong to a Cannot-Link constraint, the matrix $W$ is accordingly modified as

$$W(B_{VSC1}, B_{VSC2}) = 0, \quad (B_{VSC1}, B_{VSC2}) \in \text{Cannot-link} \quad (19)$$

The above equation suggests that the weight of the VSC-HVDC terminals should be set to zero so that the terminals can be placed in different islands to facilitate power exchange between islands.

### C. SOLVING PROCESS BASED ON SSSC

For a given AC/DC system $G_0(V, E)$, the SSSC-based solution process is as follows.

Step 1: By using WAMS information, the power flow of the power system is collected every five minutes to calculate the composite power-flow disruption.

Step 2: Construct the weighted undirected graph $G_0$ representing the AC/DC system and assign weights to every edge.

Step 3: Calculate weighted adjacency matrix $W$ and degree matrix $A$ by Eq. (20) and (21), respectively.

$$W_{ij} = \begin{cases} \dfrac{|P_{ij}|}{D_{ij}}, & (i, j) \in E_0 \\ 0, & (i, j) \notin E_0 \end{cases} \quad (20)$$

$$A_{ij} = \begin{cases} \sum_{j=1, j \neq i}^{n} W_{ij}, & i = j \\ 0, & i \neq j \end{cases} \quad (21)$$

Step 4: Detect whether a fault occurs. If it occurs, go to step 6; otherwise, go to step 5.

Step 5: Determine whether the acquisition time is met. If met, go to step 1; otherwise, go to step 4.

Step 6: Judge whether the system instability occurs. If it occurs, go to step 8; otherwise, go to the next step.

Step 7: Determine whether the instability of power system is satisfied. If satisfied, go to step 8; otherwise, go to step 6. The purpose of this step is to detect the fault information in real time to determine whether the power system has reached instability.

Step 8: When the system occurs fault and loses stability, the matrix $W$ is adjusted by Eq. (18) based on the information of coherence constraints and then update the degree matrix $A$ by Eq. (21).

Step 9: Modify the weighted adjacency matrix $W$ according to the VSC-HVDC terminal constraint and then update the degree matrix $A$.

Step 10: Calculate the normalized Laplacian matrix $L_N$ from the resulting of weighted adjacency matrix $W$ and degree matrix $A$, i.e., $L_N = A^{-1}(A-W) = A^{-1}L$.

Step 11: Calculate the eigenvectors $v_1, v_2, ..., v_k \in R^n$ that correspond to the top $k$ minimum eigenvalues of the equation $L_N v = \lambda v$.

Step 12: Build the eigenvector matrix $M = [v_1, v_2, ..., v_k] \in R^{n \times k}$, whose row vector $Rv_j \in R^k \ (j = 1, 2, ..., n)$.

Step 13: Cluster the row vectors $Rv \in R^k$ via the $k$-means algorithm, and obtain the final islanding scheme.

The flowchart of the proposed approach based on SSSC is shown in Fig. 3.

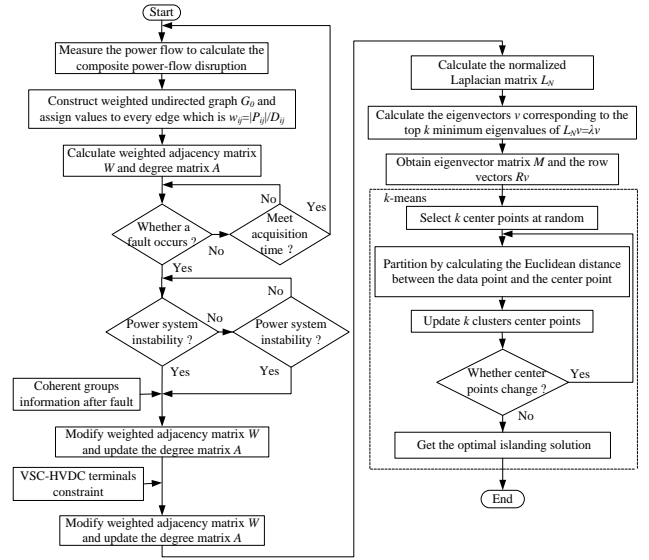

**FIGURE 3. Flowchart of the proposed controlled islanding algorithm**

### V. CASE STUDIES

In this section, the presented approach is tested on the IEEE 39-bus system and Xiamen power system. And furthermore, to reasonably evaluate the performance of the presented approach, a comparative analysis between our method and other popular methods, such as the OBDD [1], *k*-means [24] and SOM [25], has been carried out.

### A. CASE 1: IEEE 39-BUS SYSTEM

First, a modified IEEE 39-bus system is used to demonstrate the performances of the proposed method. This system is a famous test case for controlled islanding studies in the previous literature [26-28]. This system includes 10 generators, 39 buses, and 46 lines. Note that the AC line connecting buses 4 and 14 is replaced by a VSC-HVDC link. The parameters of the two-terminal VSC-HVDC system are shown in Table I. The simulations are implemented on a PC platform with 2 Intel Core dual-core CPUs (2.4 GHz) and 4 GB RAM.

TABLE I

BUS PARAMETERS OF THE VSC-HVDC SYSTEM

| Bus | $R$ (p.u.) | $X$ (p.u.) | $P_s$ (p.u.) | $Q_s$ (p.u.) | $U_{dc}$ (p.u.) |
|---|---|---|---|---|---|
| 4 | 0.002 | 0.25 | 0.561 | 0.231 | 1.000 |
| 14 | 0.002 | 0.25 | -0.563 | -0.205 | 1.000 |

The one-line diagram of this system is shown in Fig. 4.

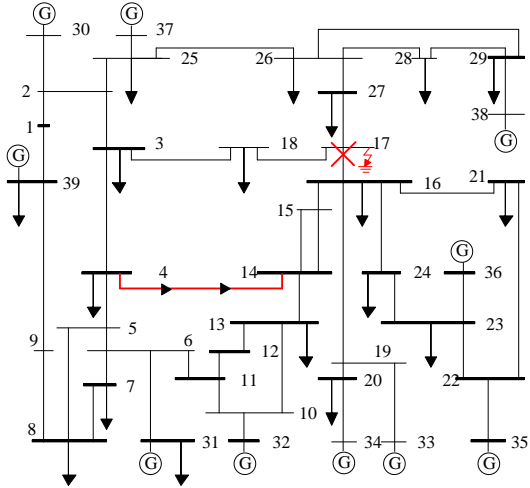

**FIGURE 4. The modified IEEE 39-bus system**

#### 1) ISLANDING SOLUTION UNDER CONVENTIONAL CONSTRAINTS

Based on power flow data from WAMS, construct weighted undirected graph $G_0$ of the system and assign values to $W_{ij}$ in the weighted adjacency matrix $W$. For convenience, only nonzero elements are listed and converted to the per unit values by dividing the same base value. The values are shown in Table II.

TABLE II

ELEMENTS MATRIX $W$ UNDER CONVENTIONAL CONSTRAINTS

| Line | $w_{ij}$ (p.u.) | Line | $w_{ij}$ (p.u.) |
|---|---|---|---|
| $L_{1-2}$ | 48.41 | $L_{14-15}$ | 2.07 |
| $L_{1-39}$ | 43.12 | $L_{15-16}$ | 32.89 |
| $L_{2-3}$ | 29.56 | $L_{16-17}$ | 25.16 |
| $L_{2-25}$ | 25.36 | $L_{16-19}$ | 23.06 |
| $L_{2-30}$ | 33.84 | $L_{16-21}$ | 29.32 |
| $L_{3-4}$ | 6.12 | $L_{16-24}$ | 7.79 |
| $L_{3-18}$ | 2.62 | $L_{17-18}$ | 26.67 |
| $L_{4-5}$ | 16.41 | $L_{17-27}$ | 0.99 |
| $L_{4-14}$ | 9.28 | $L_{19-20}$ | 12.53 |
| $L_{5-6}$ | 212.81 | $L_{19-33}$ | 43.50 |
| $L_{5-8}$ | 52.33 | $L_{20-34}$ | 28.12 |
| $L_{6-7}$ | 70.28 | $L_{21-22}$ | 52.58 |
| $L_{6-11}$ | 45.72 | $L_{22-23}$ | 5.08 |
| $L_{6-31}$ | 27.63 | $L_{22-35}$ | 45.49 |
| $L_{7-8}$ | 47.89 | $L_{23-24}$ | 18.13 |
| $L_{8-9}$ | 1.09 | $L_{23-36}$ | 20.58 |
| $L_{9-39}$ | 30.99 | $L_{25-26}$ | 3.33 |
| $L_{10-11}$ | 88.10 | $L_{25-37}$ | 23.76 |
| $L_{10-13}$ | 74.90 | $L_{26-27}$ | 20.84 |
| $L_{10-32}$ | 32.39 | $L_{26-28}$ | 4.65 |
| $L_{11-12}$ | 0.075 | $L_{26-29}$ | 5.91 |
| $L_{12-13}$ | 0.28 | $L_{28-29}$ | 26.04 |
| $L_{13-14}$ | 31.61 | $L_{29-38}$ | 53.54 |

Suppose that at time *t*=0.0s, there is a three-phase-to-ground fault occurring at line 16-17 (near bus 17), as shown in Fig. 4. The fault is cleared at *t*=0.5s. According to the generator angle deviation [24, 26], all generators are divided into three coherent groups, as shown in Table III.

TABLE III

COHERENT GROUPS OF GENERATORS

| Group number | Generator bus number |
|---|---|
| 1 | 30, 39 |
| 2 | 31, 32, 33, 34, 35, 36 |
| 3 | 37, 38 |

After obtaining the eigenvectors corresponding to the top three minimum eigenvalues of matrix $L_N$, three subsets are obtained by *k*-means clustering which is shown in Table IV. The obtained optimal islanding scheme under conventional constraints is shown in Fig. 5.

TABLE IV

CLUSTERING RESULTS UNDER CONVENTIONAL CONSTRAINTS

| Group number | Bus number |
|---|---|
| 1 | 1, 2, 3, 4, 5, 6, 7, 8, 9, 10, 11, 12, 13, 14, 25, 30, 31, 32, 37, 39 |
| 2 | 15, 16, 17, 18, 19, 20, 21, 22, 23, 24, 33, 34, 35, 36 |
| 3 | 26, 27, 28, 29, 38 |

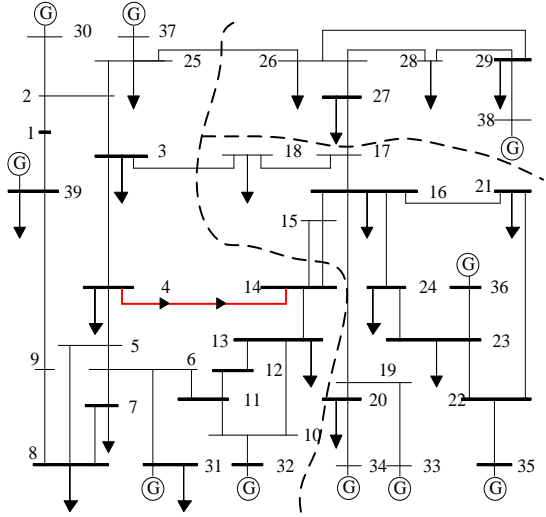

**FIGURE 5.** Islanding scheme under conventional constraints

Fig. 5 shows that the disconnected lines include $L_{3-18}$, $L_{14-15}$ (double-circuit lines), $L_{17-27}$, $L_{25-26}$, and $L_{3-18}$ in this islanding solution. It can be seen that the scheme cannot guarantee that coherent generators are divided into the same islands. For example, the generators connected to buses 37 and 38 belong to coherent generator group 3, but they are placed in different islands. Meanwhile, the VSC-HVDC terminals (buses 4 and 14) are in the same island such that the power exchanges between islands cannot be implemented.

2) ISLANDING SOLUTION UNDER COHERENCE CONSTRAINTS

Based on the pair-wise constraints of coherent generators, the elements in the $W$ matrix are modified by Eq. (18), i.e. the weight of coherent groups is equal to the infinity $w_{ij}=\infty$ while that of non-coherent groups is zero $w_{ij}=0$. The modified terms of the $W$ matrix are shown in Table V. The islanding solution is obtained by clustering the eigenvectors corresponding to the top three minimum eigenvalues of matrix $L_N$. Under this scenario, the obtained islanding solution is shown in Fig. 6.

TABLE V
MODIFIED TERMS OF MATRIX $W$ UNDER COHERENCE CONSTRAINTS

| Line | $w_{ij}$ | Line | $w_{ij}$ |
|---|---|---|---|
| $L_{30-31}$ | 0 | $L_{32-39}$ | 0 |
| $L_{30-32}$ | 0 | $L_{33-34}$ | ∞ |
| $L_{30-33}$ | 0 | $L_{33-35}$ | ∞ |
| $L_{30-34}$ | 0 | $L_{33-36}$ | ∞ |
| $L_{30-35}$ | 0 | $L_{33-37}$ | 0 |
| $L_{30-36}$ | 0 | $L_{33-38}$ | 0 |
| $L_{30-37}$ | 0 | $L_{33-39}$ | 0 |
| $L_{30-38}$ | 0 | $L_{34-35}$ | ∞ |
| $L_{30-39}$ | ∞ | $L_{34-36}$ | ∞ |
| $L_{31-32}$ | ∞ | $L_{34-37}$ | 0 |
| $L_{31-33}$ | ∞ | $L_{34-38}$ | 0 |
| $L_{31-34}$ | ∞ | $L_{34-39}$ | 0 |
| $L_{31-35}$ | ∞ | $L_{35-36}$ | ∞ |
| $L_{31-36}$ | ∞ | $L_{35-37}$ | 0 |
| $L_{31-37}$ | 0 | $L_{35-38}$ | 0 |
| $L_{31-38}$ | 0 | $L_{35-39}$ | 0 |
| $L_{31-39}$ | 0 | $L_{36-37}$ | 0 |
| $L_{32-33}$ | ∞ | $L_{36-38}$ | 0 |
| $L_{32-34}$ | ∞ | $L_{36-39}$ | 0 |
| $L_{32-35}$ | ∞ | $L_{37-38}$ | ∞ |
| $L_{32-36}$ | ∞ | $L_{37-39}$ | 0 |
| $L_{32-37}$ | 0 | $L_{38-39}$ | 0 |
| $L_{32-38}$ | 0 | | |

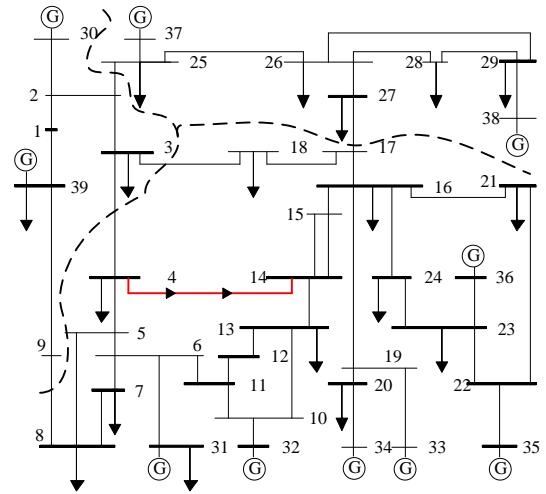

**FIGURE 6.** Islanding solution under coherence constraints

From Fig. 6, it can be observed that the disconnected lines are $L_{2-25}$, $L_{3-18}$, $L_{3-4}$, $L_{8-9}$, $L_{17-27}$. In this islanding solution, the minimum cut is 35.82, and the coherent generators are placed in the same island due to the consideration of coherence constraints, which is beneficial to the synchronization and stable operation of the islands. But similar to the solution shown in Fig. 5, all VSC-HVDC terminals are placed in the island.

3) OPTIMAL ISLANDING SCHEME

Besides the above-mentioned conventional and coherence constraints, VSC-HVDC terminal constraint is further considered to realize exchanging powers between different islands via a VSC-HVDC link. Taking into account all these constraints, an optimal islanding scheme is obtained, as shown in Fig. 7.

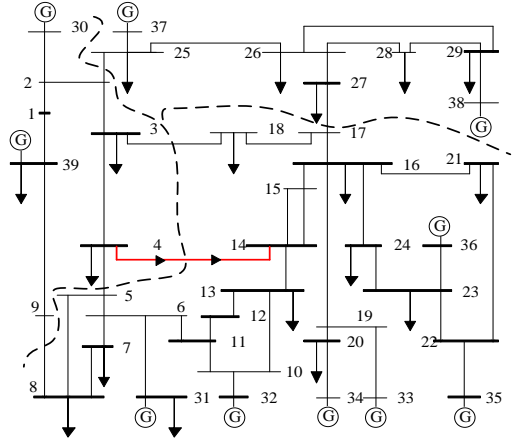

**FIGURE 7.** Optimal islanding scheme

Fig. 7 demonstrates that the disconnected lines are $L_{2-25}$, $L_{3-18}$, $L_{4-5}$, $L_{8-9}$, $L_{17-27}$ in this islanding scheme. In this way, the coherent generators are divided into the same island and the terminals of a VSC-HVDC link are located in different islands. And thereby, the power exchanges between islands can be performed through VSC-HVDC links to reduce the generation-load imbalance, which is helpful to accelerate the follow-up power system restoration. The total computing time is 0.016 s, which satisfies the real-time requirements in practical applications.

The obtained splitting schemes using different methods are shown in Table VI.

TABLE VI

RESULTS OF THE IEEE 39-BUS SYSTEM USING DIFFERENT METHODS

| Method | Disconnected lines | $\sum w_{ij}$ (p.u.) | Running time(s) |
|---|---|---|---|
| SSSC | $L_{2-25}$, $L_{3-18}$, $L_{4-5}$, $L_{4-14}$, $L_{8-9}$, $L_{17-27}$ | 55.39 | 0.016 |
| OBDD | $L_{2-25}$, $L_{3-18}$, $L_{4-5}$, $L_{4-14}$, $L_{8-9}$, $L_{17-27}$ | 55.39 | 2.231 |
|  | $L_{2-25}$, $L_{3-18}$, $L_{4-5}$, $L_{4-14}$, $L_{9-39}$, $L_{17-27}$ | 85.29 |  |
|  | $L_{2-25}$, $L_{3-18}$, $L_{4-5}$, $L_{4-14}$, $L_{8-9}$, $L_{17-18}$, $L_{16-17}$ | 106.23 |  |
| SOM | $L_{2-25}$, $L_{3-18}$, $L_{4-5}$, $L_{4-14}$, $L_{8-9}$, $L_{17-18}$, $L_{16-17}$ | 106.23 | 1.807 |
| k-means | $L_{1-2}$, $L_{2-30}$, $L_{3-18}$, $L_{4-5}$, $L_{4-14}$, $L_{8-9}$, $L_{17-27}$ | 112.28 | 0.167 |

From Table VI, it can be seen that the proposed approach manages to directly find the optimal solution and it is superior to other alternatives. 1) Regarding the OBDD: the OBDD can give all possible solutions, while our approach can directly determine the optimal solution with the minimum cut $\sum w_{ij}$. The reason for this is that the OBDD is based on a graphical representation of Boolean functions, while our approach can directly yield the optimal solution by transforming the original problem into a graph segmentation issue. 2) For the k-means: the $L_{1-2}$ and $L_{2-30}$ should be disconnected in its solution, which will inevitably cause the generator connected to bus 30 to be tripped and increase the splitting costs. The reason is that the k-means, as a "hard clustering" algorithm, is prone to local minima by simply clustering the data [42]. On the contrary, the SSSC can obtain the global optimal solution and avoid the formation of isolated nodes due to the normalized cut criterion. 3) Regarding the SOM: there is no isolated node in its scheme, but the $\sum w_{ij}$ is significantly greater than that of the SSSC.

The computational efficiency is another important evaluation indicator for this problem since the islanding system control is a fast process. As can be seen from Table VI, the computational efficiency of our approach is far superior to that of other methods. Specifically, the computational times of the OBDD, SOM and k-means are respectively 2.231 s, 1.807s and 0.167s, while our approach only requires 0.016 s. Therefore, a conclusion can be drawn that the presented algorithm is an effective tool to solve the controlled islanding problem.

### B. CASE 2: XIAMEN POWER SYSTEM

In order to further verify the applicability of the proposed method to real-world systems, the Xiamen power system is utilized as the testing system. As illustrated in Fig. 8, this system consists of 8 generators, 28 buses, 36 AC lines, and 1 VSC-HVDC link. Specifically speaking, two generators are connected to buses 4 and 24, and four equivalent generators are respectively connected to buses 1, 2, 14, and 17. The used VSC-HVDC parameters are the same as those in case 1.

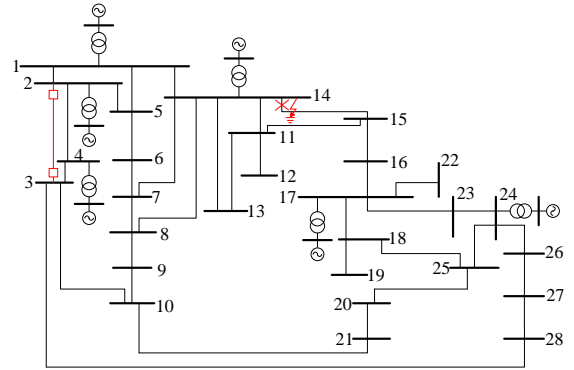

**FIGURE 8.** One-line diagram of the Xiamen power system

Simulation settings: The system begins to operate at 0 s and the VSC-HVDC link is turned on at $t = 15$ s. At 16s, a three-phase-to-ground fault occurs on the line $L_{14-15}$ (near bus 14) as shown in Fig. 8, and the fault is cleared at $t = 17$s. The entire simulation lasts for 50 s. Based on the power flow data from the WAMS, the composite power-flow disruption and the matrix $W$ can be obtained. For ease of description, only nonzero elements of matrix $W$ are listed in Table VII.

TABLE VII

ELEMENTS OF MATRIX $W$ UNDER CONVENTIONAL CONSTRAINTS

| Line | $w_{ij}$ (p.u.) | Line | $w_{ij}$ (p.u.) |
|---|---|---|---|
| $L_{1-2}$ | 23.65 | $L_{11-14}$ | 16.53 |
| $L_{1-5}$ | 53.4 | $L_{11-15}$ | 11.28 |
| $L_{1-14}$ | 67.77 | $L_{13-14}$ | 6.48 |
| $L_{2-4}$ | 1.49 | $L_{14-15}$ | 9.08 |
| $L_{2-5}$ | 22.62 | $L_{15-16}$ | 15.04 |
| $L_{3-4}$ | 74.64 | $L_{16-17}$ | 27.89 |
| $L_{3-10}$ | 33.03 | $L_{17-18}$ | 56.0 |
| $L_{3-28}$ | 22.26 | $L_{17-22}$ | 10.7 |
| $L_{5-6}$ | 50.25 | $L_{17-23}$ | 20.32 |
| $L_{6-7}$ | 34.16 | $L_{18-19}$ | 9.63 |
| $L_{7-8}$ | 21.35 | $L_{18-25}$ | 24.93 |
| $L_{7-14}$ | 25.76 | $L_{20-21}$ | 14.21 |
| $L_{8-9}$ | 23.46 | $L_{20-25}$ | 33.52 |
| $L_{8-14}$ | 37.44 | $L_{23-24}$ | 1.04 |
| $L_{9-10}$ | 5.5 | $L_{24-25}$ | 24.72 |
| $L_{10-21}$ | 45.8 | $L_{24-26}$ | 51.62 |
| $L_{11-12}$ | 21.41 | $L_{26-27}$ | 22.68 |
| $L_{11-13}$ | 6.47 | $L_{27-28}$ | 3.42 |

In this case, the angle deviations between generators are obviously different after fault clearance. According to the coherent grouping approach used in case 1, the generators are divided into two groups: two inside generators and four equivalent generators, and thereby the system is split into two asynchronous islands.

Besides conventional constraints, the coherence and VSC-HVDC terminal constraints are employed to modify the elements of the matrix $W$, in which the modified elements are shown in Table VIII.

Table VIII

MODIFIED TERMS OF $W$ MATRIX UNDER COHERENCE CONSTRAINTS AND VSC-HVDC TERMINAL CONSTRAINT

| Line | $w_{ij}$ | Line | $w_{ij}$ |
|---|---|---|---|
| $L_{1-2}$ | ∞ | $L_{2-17}$ | ∞ |
| $L_{1-4}$ | 0 | $L_{2-24}$ | 0 |
| $L_{1-14}$ | ∞ | $L_{4-14}$ | 0 |
| $L_{1-17}$ | ∞ | $L_{4-17}$ | 0 |
| $L_{1-24}$ | 0 | $L_{4-24}$ | ∞ |
| $L_{2-3}$ | 0 | $L_{14-17}$ | ∞ |
| $L_{2-4}$ | 0 | $L_{14-24}$ | 0 |
| $L_{2-14}$ | ∞ | $L_{17-24}$ | 0 |

The eigenvectors that correspond to the top two minimum eigenvalues of matrix $L_N$ are obtained. After clustering, the optimal islanding solution is shown in Fig. 9.

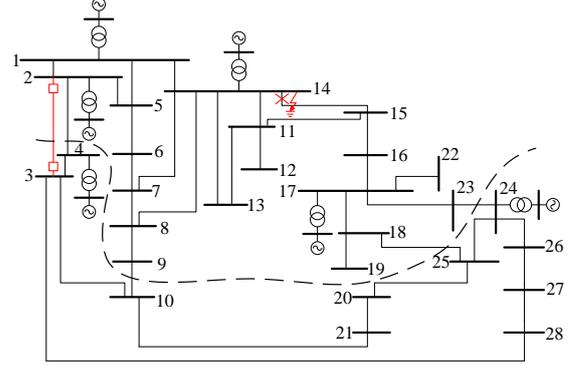

**FIGURE 9.** Optimal islanding scheme

From Fig. 9, it can be observed that the disconnected lines are $L_{2-4}$, $L_{9-10}$, $L_{18-25}$, $L_{23-24}$. And then, the comparative tests have been executed and the obtained results are shown in Table IX.

TABLE IX

RESULTS OF THE XIAMEN SYSTEM USING DIFFERENT METHODS

| Method | Disconnected lines | $\sum w_{ij}$ (p.u.) | Running time(s) |
|---|---|---|---|
| SSSC | $L_{2-4}$, $L_{9-10}$, $L_{18-25}$, $L_{23-24}$ | 32.96 | 0.012 |
| OBDD | $L_{2-4}$, $L_{9-10}$, $L_{18-25}$, $L_{23-24}$ | 32.96 | 2.125 |
|  | $L_{2-4}$, $L_{8-9}$, $L_{18-25}$, $L_{23-24}$ | 50.92 |  |
|  | $L_{2-4}$, $L_{9-10}$, $L_{18-25}$, $L_{17-23}$ | 52.24 |  |
| SOM | $L_{2-4}$, $L_{9-10}$, $L_{18-25}$, $L_{17-23}$ | 52.24 | 0.995 |
| k-means | $L_{2-4}$, $L_{8-9}$, $L_{20-21}$, $L_{23-24}$, $L_{24-25}$ | 64.37 | 0.159 |

From Table IX, it can be observed that the SSSC outperforms the other methods in this case. First, the $\sum w_{ij}$ of the SSSC is significantly less than that of the SOM and k-means. Second, the computational efficiency of the SSSC is obviously better than that of other methods, which suggests that the SSSC can better meet the real-time requirements in practical applications. As a result, the effectiveness and superiority of the SSSC in real systems can be verified. In order to examine the effects of the obtained splitting scheme using the proposed approach, simulation analysis has been performed under the RT-LAB simulation platform. The voltage amplitudes of all generators during the splitting process are shown in Figs. 10-12.

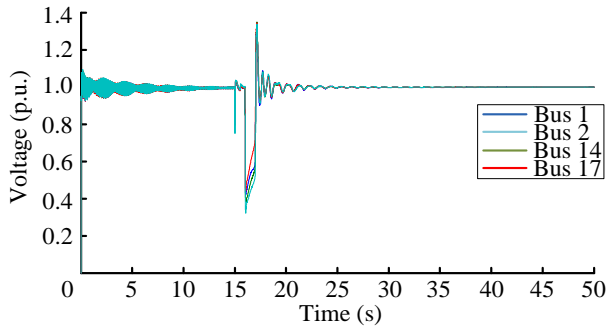

**FIGURE 10.** Bus voltages of the equivalent generators

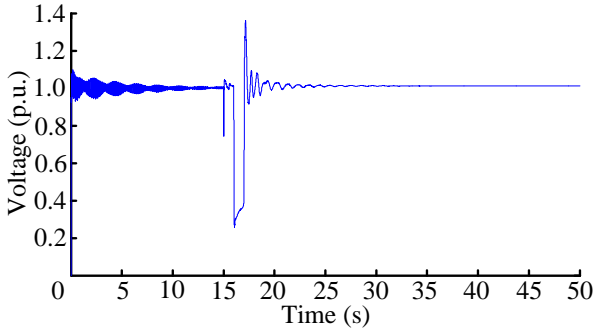

**FIGURE 11.** Bus voltages of generators 4

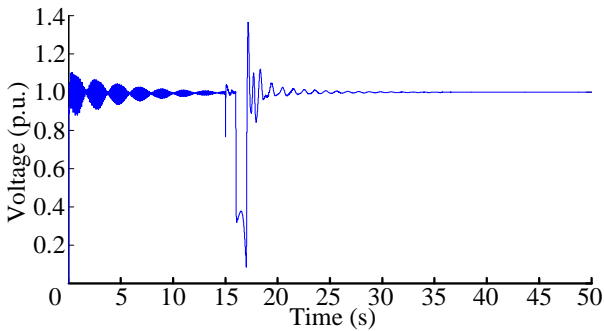

**FIGURE 12.** Bus voltages of generators 24

From Figs. 10-12, it can be seen that when the generators start, the generator bus voltages will produce some fluctuations and gradually tend to be stable. After a severe fault occurs at 16s, the system begins to lose stability. Meanwhile, the generator bus voltages show violent fluctuations. After fault clearance, the voltage fluctuations gradually decrease and eventually tend to be stable. The reason for this phenomenon is that after the fault is cleared the proposed method will be triggered at once, and thereby the oscillations quickly cease and the system regains stable due to power exchanges between islands through the VSC-HVDC link. Specifically speaking, according to the obtained optimal islanding scheme the system is split into two islands, where the coherent generators are divided into the same island due to the coherence constraints and the two VSC-HVDC terminals are placed in different islands because of the VSC-HVDC terminal constraints. In this scenario, the two islands are connected by the VSC-HVDC link. In this way, the entire system quickly restores stability because emergency power supports are provided for the islands on both sides under fault conditions with the full use of the powerful DC power modulation ability of VSC-HVDC. This result not only validates the applicability of the proposed method to the actual Xiamen power system but also further confirms the role of VSC-HVDC for enhancing the safety and stability of the AC/DC system.

## VI. CONCLUSIONS

The recently developed VSC-HVDC technology is becoming a popular choice for the delivery of bulk power, and emergency control issues on power systems need to be paid more attention due to increasing uncertainties in system operation under the growing penetration of renewable energies. To make full use of the powerful flexible DC power modulation capability, this paper presents a new controlled islanding model for a hybrid AC/VSC-HVDC grid. To solve this model, a solution method based on SSSC is presented by integrating the constraints of generator coherences and VSC-HVDC terminals besides conventional ones. By doing so, the optimal islanding solution problem can be transformed into a graph segmentation issue which is readily solvable by using the constrained spectral clustering. Studies carried out for IEEE 39-bus system and the real-world Xiamen power system reveal that our approach is a good candidate for addressing controlled islanding problem. The proposed algorithm manages to find an optimal islanding scheme to enable a VSC-HVDC link to connect different islands such that the generation-load imbalance can be significantly reduced due to power exchanges between islands. More importantly, the superiority of our approach to other popular alternative methods, including the OBDD, SOM and $k$-means clustering, has also been verified. The proposed approach might find potential applications in real-time system splitting and restoration for a WAPaC system.

The research on controlled islanding for a hybrid AC/VSC-HVDC system is still in relatively early stages of development. Future work will focus on developing more realistic modeling techniques and decision analysis approaches [43] to consider more complex fault scenarios such as one or more converter blocking failures. Besides, the control and coordination problem of VSC-based multi-terminal DC (VSC-MTDC) systems in the event of system failure is another interesting topic for future research.

## REFERENCES
[1] K. Sun, D. Z. Zheng, and Lu Q, "Splitting strategies for islanding operation of large-scale power systems using OBDD-based methods," *IEEE Trans. Power Syst.*, vol. 18, no. 2, pp. 912-923, 2003.
[2] S. S. Ahmed, N. C. Sarker, A. B. Khairuddin, M. R. B. A. Ghani, and H. Ahmad, "A scheme for controlled islanding to prevent subsequent blackout," *IEEE Trans. Power Syst.*, vol. 18, no. 1, pp. 136-143, 2003.

[3] N. Senroy, and G. T. Heydt, "A conceptual framework for the controlled islanding of interconnected power systems," *IEEE Trans. Power Syst.*, vol. 21, no. 2, pp. 1005-1006, 2006.
[4] S. P. Wang, A. Chen, and C. W. Liu, "Efficient splitting simulation for blackout analysis," *IEEE Trans. Power Syst.*, vol. 30, no. 4, pp. 1775-1783, 2015.
[5] A. Kyriacou, P. Demetriou, C. Panayiotou, and E. Kyriakides, "Controlled islanding solution for large-scale power systems," *IEEE Trans. Power Syst.*, vol. 33, no. 2, pp. 1591-1602, 2018.
[6] Y. Li, B. Feng, G. Li, J. Qi, D. Zhao, and Y. Mu, "Optimal distributed generation planning in active distribution networks considering integration of energy storage," *Appl. Energy*, vol. 210, pp. 1073-1081, 2018.
[7] Y. Li, Z. Yang, G. Li, D. Zhao, and W. Tian, "Optimal scheduling of an isolated microgrid with battery storage considering load and renewable generation uncertainties," *IEEE Trans. Ind. Electron.*, to be published. DOI: 10.1109/TIE.2018.2840498
[8] P. Jin, Y. Li, and G. Li, "Optimized hierarchical power oscillations control for distributed generation under unbalanced conditions," *Appl. Energ.*, vol. 194, no. 343-352, 2016.
[9] P. Y. Kong, and G. K. Karagiannidis, "Charging schemes for plug-in hybrid electric vehicles in smart grid: a survey," *IEEE Access*, vol. 4, no.99, pp. 6846-6875, 2016.
[10] Y. Li, Z. Yang, and G. Li G, "Optimal scheduling of isolated microgrid with an electric vehicle battery swapping station in multi-stakeholder scenarios: A bi-level programming approach via real-time pricing," *Appl. Energ.*, vol. 232, pp. 54-68, 2018.
[11] Y. Li, and Z. Yang, "Application of EOS-ELM with binary Jaya-based feature selection to real-time transient stability assessment using PMU data," *IEEE Access*, vol. 5, pp. 23092-23101, 2017.
[12] C. Chen, J. Wang, and D. Ton, "Modernizing distribution system restoration to achieve grid resiliency against extreme weather events: An integrated solution," *Proceedings of the IEEE*, vol. 105, no. 7, pp. 1267-1288, 2017.
[13] Z. Huang, C. Wang, and T. Zhu, "Cascading failures in smart grid: Joint effect of load propagation and interdependence," *IEEE Access*, vol. 3, pp. 2520-2530, 2015.
[14] B. A. Carreras, D. E. Newman, and I. Dobson, "North American blackout time series statistics and implications for blackout risk," *IEEE Trans. Power Syst.*, vol. 31, no. 6, pp. 4406-4414, 2016.
[15] C. G. Wang, B. H. Zhang, and Z. G. Hao, "A novel real-time searching method for power system splitting boundary," *IEEE Trans. Power Syst.*, vol. 25, no. 4, pp. 1902-1909, 2010.
[16] M. Golari, N. Fan, and J. Wang, "Two-stage stochastic optimal islanding operations under severe multiple contingencies in power grids," *Electr. Power Syst. Res.*, vol. 114, pp. 68-77, 2014.
[17] X. Liu, J. M. Kennedy, and D. M. Laverty, "Wide-area phase-angle measurements for islanding detection—An adaptive nonlinear approach," *IEEE Trans. Power Syst.*, vol. 31, no. 4, pp. 1901-1911, 2016.
[18] J. Li, C. C. Liu, and K. P. Schneider, "Controlled partitioning of a power network considering real and reactive power balance," *IEEE Trans. Smart grid*, vol. 1, no. 3, pp. 261-269, 2010.
[19] H. You, V. Vittal, and X. Wang, "Slow coherency-based islanding," *IEEE Trans. Power Syst.*, vol. 19, no. 1, pp. 483-491, 2004.
[20] B. Yang, V. Vittal, and G. T. Heydt, "Slow-coherency-based controlled islanding—a demonstration of the approach on the August 14, 2003 blackout scenario," *IEEE Trans. Power Syst.*, vol.21, no.4, pp.1840-1847, 2006.
[21] G. Xu, and V. Vittal, "Slow coherency based cutset determination algorithm for large power systems," *IEEE Trans. Power Syst.*, vol. 25, no. 2, pp. 877-884, 2010.
[22] Q. Zhao, K. Sun, D. Zheng, J. Ma, and Q. Lu, "A study of system splitting strategies for island operation of power system: a two-phase method based on OBDDs," *IEEE Trans. Power Syst.*, vol. 18, no.4, pp. 1556-1565, 2003.
[23] S. Kai, Z. Dazhong, and L. Qiang, "A simulation study of OBDD-based proper splitting strategies for power systems under consideration of transient stability," *IEEE Trans. Power Syst.*, vol. 20, no. 1, pp. 389-399, 2005.
[24] M. Imhof, O. Valgaev, and G. Andersson, "Controlled islanding using VSC-HVDC links to reduce load shedding," *IEEE Power Tech*, pp. 1-6, Eindhoven, Jun. 2015.
[25] M. Mahdi, and I. Genc, "Defensive islanding using self-organizing maps neural networks and hierarchical clustering," *IEEE Power Tech.*, pp. 1-5, Eindhoven, 2015.
[26] L. Ding L, F. M. Gonzalez-Longatt, and P. Wall, "Two-step spectral clustering controlled islanding algorithm," *IEEE Trans. Power Syst.*, vol. 28, pp. 75-84, 2013.
[27] J. Quirós-Tortós, R. Sánchez-García, and J. Brodzki, "Constrained spectral clustering-based methodology for intentional controlled islanding of large-scale power systems," *IET Gener. Transm. Dis.*, vol. 9, no. 1, pp. 31-42, 2014.
[28] R. J. Sánchez-García, M. Fennelly, and S. Norris, "Hierarchical spectral clustering of power grids," *IEEE Trans. Power Syst.*, vol. 29, no. 5, pp. 2229-2237, 2014.
[29] W. Chen, and G. Feng, "Spectral clustering: A semi-supervised approach," *Neurocomputing*, vol. 77, no. 1, pp. 229-242, 2012.
[30] Y. Li, Y. Li, G. Li, D. Zhao, and C. Chen, "Two-stage multi-objective OPF for AC/DC grids with VSC-HVDC: Incorporating decisions analysis into optimization process," *Energy*, vol. 147, pp. 286–296, 2018.
[31] X. Zhang, "Multiterminal voltage-sourced converter-based HVDC models for power flow analysis," *IEEE Trans. Power Syst.*, vol. 19, pp. 1877–1884, 2004.
[32] M. A. Abdelwahed, and E. F. El-Saadany, "Power sharing control strategy of multi-terminal VSC-HVDC transmission systems utilizing adaptive voltage droop," *IEEE Trans. Sustain. Energ.*, vol. 8, no. 2, pp. 605-615, 2017.
[33] N. Flourentzou, V. G. Agelidis, and G. D. Demetriades, "VSC-Based HVDC power transmission systems: an overview," *IEEE T. Power Electr.*, vol. 24, no. 3, pp. 592-602, 2009.
[34] D. V. Hertem, M. Ghandhari. "Multi-terminal VSC HVDC for the European supergrid: obstacles," *Renewable and sustainable energy reviews*, vol. 14, no. 9, pp. 3156-3163, 2010.
[35] L. Zhang, L. Harnefors, and H. P. Nee, "Modeling and control of VSC-HVDC links connected to island systems," *IEEE Trans. Power Syst.*, vol. 26, no. 2, pp. 783-793, 2011.
[36] X. Gu, Y. Li, and J. Jia, "Feature selection for transient stability assessment based on kernelized fuzzy rough sets and memetic algorithm," *Int. J. Electr. Power Energy Syst.*, vol. 64, no. 64, pp. 664-670, 2015.
[37] V. Terzija, G. Valverde, D. Cai, P. Regulski, V. Madani, J. Fitch, S. Skok, M. M. Begovic, and A. Phadke, "Wide-area monitoring, protection, and control of future electric power networks," *Proc. IEEE*, vol. 99, no. 1, pp. 80–93, Jan. 2011.
[38] Y. Li, G. Li, Z. Wang, Z. Han, and X. Bai, "A multifeature fusion approach for power system transient stability assessment using PMU data," *Math. Problems Eng.*, vol. 2015, 2015, Art. no. 786396.
[39] L. Zhang, L. Harnefors, and H. P. Nee, "Interconnection of two very weak AC systems by VSC-HVDC links using power-synchronization control," *IEEE Trans. Power Syst.*, vol. 26, no. 1, pp. 344-355, 2011.
[40] J. Beerten, S. Cole S, and R. Belmans, "Generalized steady-state VSC MTDC model for sequential AC/DC power flow algorithms," *IEEE Trans. Power Syst.*, vol.27, no.2, pp.821-829, 2012.
[41] E. Cotilla-Sanchez, P. D. H. Hines, and C. Barrows, "Multi-attribute partitioning of power networks based on electrical distance," *IEEE Trans. Power Syst.*, vol. 28, no. 4, pp. 4979-4987, 2013.
[42] A. Rodriguez, and A. Laio, "Clustering by fast search and find of density peaks," *Science*, vol. 344, no. 6191, pp. 1492-1496, 2014.
[43] Y. Li, J. Wang, and D. Zhao, "A two-stage approach for combined heat and power economic emission dispatch: Combining multi-objective optimization with integrated decision making," *Energy*, vol. 162, pp. 237-254, 2018.